\documentclass[runningheads]{llncs}
\usepackage{graphicx}
\usepackage{hyperref}  
\usepackage{adjustbox}
\usepackage{subfig}

\usepackage{xcolor}

\begin{document}
\title{Exploring the Universe with SNAD: Anomaly Detection in Astronomy}
%
%
\author{Alina A. Volnova\inst{1}\orcidID{0000-0003-3554-1037} 
\and Patrick D. Aleo\inst{2,3}\orcidID{0000-0002-6298-1663}
\and Anastasia Lavrukhina\inst{4}
\and Etienne Russeil\inst{5}\orcidID{0000-0001-9923-2407} 
\and Timofey Semenikhin\inst{4,6}
\and Emmanuel Gangler\inst{5}
\and Emille E. O. Ishida\inst{5} \orcidID{0000-0002-0406-076X} 
\and Matwey V. Kornilov\inst{6,7}\orcidID{0000-0002-5193-9806}
\and Vladimir Korolev\orcidID{0009-0008-7691-6142}
\and Konstantin Malanchev\inst{2,6} \orcidID{0000-0001-7179-7406} 
\and Maria V. Pruzhinskaya\inst{5}\orcidID{0000-0001-7178-0823}    
\and Sreevarsha Sreejith\inst{8}\orcidID{0000-0002-6423-1348} }
\authorrunning{A. Volnova et al.}
%
\institute{Space Research Institute of the Russian Academy of Sciences, Profsoyuznaya 84/32, Moscow 117997, Russia \email{alinusss@gmail.com}
\and Department of Astronomy, University of Illinois at Urbana-Champaign, 1002 W. Green St., IL 61801, USA
\and Center for AstroPhysical Surveys, National Center for Supercomputing Applications, Urbana, IL, 61801, USA
\and Faculty of Space Research, Lomonosov Moscow State University, Leninsky Gori 1 bld. 52, Moscow 119234, Russia
\and Universit\'e Clermont Auvergne, CNRS/IN2P3, LPC, F-63000 Clermont-Ferrand, France
\and Sternberg Astronomical Institute, Lomonosov Moscow State University, Universitetsky 13, Moscow 119234, Russia 
\and National Research University Higher School of Economics, 21/4 Staraya Basmannaya Ulitsa, Moscow 105066, Russia
\and  Physics department, University of Surrey, Stag Hill, Guildford GU2 7XH, UK}

\maketitle              
\begin{abstract}

SNAD is an international project with a primary focus on detecting astronomical anomalies within large-scale surveys, using active learning and other machine learning algorithms. The work carried out by SNAD not only contributes to the discovery and classification of various astronomical phenomena but also enhances our understanding and implementation of machine learning techniques within the field of astrophysics. This paper provides a review of the SNAD project and summarizes the advancements and achievements made by the team over several years. 

\end{abstract}
\section{Introduction}

In modern astronomy, discoveries of new objects are based on a huge flow of data coming from all-sky surveys (e.g., the Sloan Digital Sky Survey, SDSS \cite{SDSS}, the Zwicky Transient Facility, ZTF \cite{ZTF}, the Vera Rubin Observatory Legacy Survey of Space and Time, LSST \cite{LSST}). The terabytes of data generated every night contain information, allowing the discovery of several hundred transients per year. Undoubtedly, the human resource is limited, and each scientist should choose from a variety of discovered objects those that are of the greatest interest to him. However, this is not a trivial task, given the number of transients to choose from. In addition, the total mass of discovered objects may contain rare or not yet known phenomena. Thus, we are faced with the problem of classifying a large number of objects, and specifically in this case: the search for anomalies or outliers.

It is natural to expect that such large volumes of data that are generated in astronomy today require machine learning (ML) methods for processing. Despite the fact that ML has become an integral part of data analysis in almost all areas of science in recent years, astronomy has benefited from it only recently, starting from solving problems in classification and regression (see, e.g., \cite{2019MNRAS.483....2I}). In the light of the search for the most interesting objects among all detected ones anomaly detection (AD) algorithms have a wide field of application: would it be search for galaxies with anomalous spectra \cite{2017MNRAS.465.4530B}, transient with extraordinary light curves \cite{2019OSC}, or unusual variable stars \cite{Malanchev2021}. 

However, most of those algorithms are based on statistical models, and the anomaly is defined as an object that does not fit the model. Observational defects lead to multiple identifications of 
non-physical events as anomalous, giving the researcher hundreds of candidates for further investigations. To decrease the amount of ``not interesting'' objects and to detect the astrophysical anomalies AD algorithms need some advisory from human in which object should be considered as anomalous, i.e., active learning (AL) is required. AL is a subclass of ML algorithms where the user may adjust the model by interactively setting scores to the objects which are suspicious for the machine. In this 
learning paradigm, the user can 
select which type of object should be considered as anomalous, extracting some rare classes of objects from the bulk of huge sky-survey data (see, e.g., \cite{2019MNRAS.483....2I,2020MNRAS.491.1554W}). 

Given to its important impact in the future of astronomical discoveries with modern data sets, the problem of anomaly detection have already been explored in the literature. This includes recent studies using data streams \cite{2023AJ....166..151P},  deep \cite{2023MLS&T...4b5013C}, generative \cite{2021MNRAS.508.2946S} and active \cite{IshidaAAD,2023arXiv230908660E} methods. Nevertheless, the incidence of false positives continues to be an important bottle neck to be overcome if we intend to take full advantadge of modern astronomical data. We describe here one of such attempts, which has been consistently focusing on anomaly detection for astronomy in the last 5 years.

This paper is an overview of the SNAD project -- an  international group of researchers (including astronomers, physicists, data scientists and mathematicians)  which focus on developing new AL algorithms for the search for anomalous objects in large data-sets of all-sky astronomical surveys. In Section \ref{sec:snad} we provide the description and main goals of the project and the team, and briefly summarise data sources and used techniques. In Section \ref{results} the most prominent results of the project are presented. Section \ref{sec:public} describes some auxiliary products developed by the team, which are now publicly available. Section \ref{future} gives conclusive remarks for future prospects of the project.


\section{What is the SNAD project?}
\label{sec:snad}
\subsection{Goals and objectives}

The goal of the SNAD\footnote{\url{https://snad.space}} project is to develop a pipeline where human expertise and modern machine learning techniques can complement each other in the task of identifying unusual astronomical objects mostly by their photometrical features. The team concentrates on the search for unusual, rare or yet unknown objects in the sets of photometric light curves (LCs) by combining  different AL AD algorithms with the additional information provided by the expert in order to label a significant fraction of  the most obvious outliers and choose those which are true astronomical anomalies. 
Enabling reliable anomaly/outlier detection based solely on photometric observations is one of the fundamental puzzles to be solved before we can convert the full potential of large-scale surveys into scientific results. This project represents an effective strategy to guarantee we shall not overlook exciting new science hidden in the data we fought so hard to acquire.

\subsection{Team members and expertise}
The SNAD team is composed of young researchers, each offering a unique set of skills and experiences that are utilized within the project. All team members are involved in various stages of the anomaly detection process, from feature engineering to expert analysis.

In addition to their collective expertise in anomaly detection, each team member also contributes their unique knowledge from different fields: gamma-ray bursts (Alina Volnova), supernovae (Maria Pruzhinskaya), fast transients (Anastasia Lavrukhina), accretion flows in astrophysics (Konstantin Malanchev), adapting machine learning algorithms (Emille E. O. Ishida), astronomical site characterization (Matwey Kornilov), time-domain astronomy and photometric classification (Partick Aleo), astronomical image analysis and dwarf galaxy detection (Sreevarsha Sreejith), real-bogus classification (Timofey Semenikhin), aeronautical science (Vladimir Korolev), symbolic regression for light curve analysis (Etienne Russeil), observational cosmology (Emmanuel Gangler).

Over the years, several other researchers, including Florian Mondon, Anastasia Malancheva, Alexandra Novinskaya, and Anastasiya Voloshina have also contributed their skills and knowledge to the project. 

SNAD activities are designed around annual meetings whose format was inspired by other initiatives which focused on boosting innovation and creativity (e.g. the silicon valley model \cite{silicon} and the Cosmostatistics Initiative\footnote{\url{cosmostatistics-initiative.org/}} ). We also invite\footnote{\url{https://snad.space/\#contact}} other researchers and students of different levels for collaboration in the field of astronomical anomalies detection. The resulting environment is one of the key components which play an important role in the development of products described in this work.

\subsection{Data sources and techniques}
\label{subsec:data_models}

The first data-set used for our research was extracted from the Open Supernova Catalog\footnote{\url{https://github.com/astrocatalogs}} (OSC; \cite{OSC2017}). It contains all publicly available data on all SNe from a dozen of catalogues, including, in different cases, multi-colour LCs, spectra, redshift estimations and classification. We extracted 1999 LCs with the number of data points enough to fit the LC within some particular interval of time relative to the maximum, and presented in 3 different photometric pass-bands ($gri, g'r'i'$ or $BRI$). These 1999 LCs allowed us to test our main ideas by finding several peculiar and super-luminous SNe, along with a few dozens of miss-classified objects \cite{2019OSC}.

The results of the first try encouraged us to use more numerous data-set of photometric LCs, and the Zwicky Transient Facility (ZTF) survey results were chosen. The ZTF survey started on March 2018 and during its initial phase has observed around a billion objects \cite{ZTF}. Each object is represented with a bunch of LCs in filters $zg, zr, zi$ with a cadence -- on average $\sim$ 1 day for the Galactic plane and $\sim$ 3 days
for the Northern-equatorial sky. To minimize the affection of different cadences on the results we selected 3 different fields: 1 in the M31 galaxy, 1 in the Galactic plane, and 1 far above the Galactic disk and analysed data from the first 9.4 months of the ZTF survey, between 17 March and 31 December 2018. The total amount of objects with at least 100 data-point in the LC is 2.25 millions. The results of the search for anomalies in this set are described in Section \ref{results}.

Investigating all data described above we experimented with a series of different anomaly detection algorithms. Our goal was to quantify their effectiveness in identifying scientifically interesting anomalous objects within large data sets. Among others, we used Isolation Forest (IF) \cite{ISOForest},  Local Outlier Factor \cite{Breunig_etal2000}, Gaussian Mixture Models \cite{MCLA2000} and one-class Support Vector Machines \cite{schonesvm}. These were applied to small \cite{2019OSC} as well as large \cite{Malanchev2021} data sets with encouraging results. 

However, we noticed that frequently, the majority of objects with high anomaly score represent non-astrophysical artificial effects. These were most times border effects, glitches in the detector, bad CCDs, cosmic rays, unexpected telescope movements, satellites, etc. This discovery was important, and generated an additional line of investigation, described in Section \ref{subsec:artefacts}. Nevertheless, the need of an adaptive algorithm, which could allow the user to define which type of anomaly was interesting became increasingly more evident. 

We started experimenting with adaptive learning techniques by using the Active Anomaly Discovery algorithm (AAD) \cite{Das2017}, which uses a human-in-the-loop strategy to apply a series of sequential modifications to a traditional IF. By downgrading decision paths which disagree with the expert's definition of what is an interesting anomaly the algorithm allows the construction of a personalized AD model. We showed that this strategy is effective in small \cite{IshidaAAD} and large \cite{Pruzhinskaya2023} data sets. The successful experiences with AAD motivated many of the results described below. 



\section{Key Contributions and Discoveries}
\label{results}

\subsection{Supernova catalog}
The SNAD team discovered potential, unreported supernovae --- powerful and bright explosion of a star, indicating the final stage of its evolution~\cite{2017suex.book.....B} --- within the ZTF DRs during a non-targeted anomaly detection search \cite{Malanchev2021}. This led to the inception of a new experiment aimed at developing specialized machine learning models. The AAD algorithm (Section~\ref{subsec:data_models}) and the SNAD Transient Miner (Section~\ref{miner}) have been trained and refined based on our long-term experience with supernovae. Applying these algorithms, the SNAD team identified 144 new supernova candidates hidden within the vast photometric data of the ZTF survey.


Each of these candidates has been thoroughly inspected and validated by our domain experts. Detailed information about the candidates has been made publicly available in the SNAD supernova catalog\footnote{\url{https://snad.space/catalog/}}.

Furthermore, these supernova candidates have also been reported to the Transient Name Server\footnote{\url{https://www.wis-tns.org/}} (TNS). The TNS is the official resource for announcing new astronomical transients, and our reports ensure that these candidates are available for further study by the international astronomical community. With this dual approach of cataloguing and reporting, we aim to foster collaboration and further our understanding of supernovae.

\subsection{Superluminous supernova candidates}
\label{SLSN}

Among the supernova candidates discovered using the AAD algorithm (Section~\ref{subsec:data_models}) we report four objects: SNAD120, SNAD121, SNAD160, SNAD187, which could belong to the superluminous supernovae class. They display an unusually broad LC when compared to the Nugent's models\footnote{\url{https://c3.lbl.gov/nugent/nugent_templates.html}} and multiple redshift estimations indicate their high absolute brightness~\cite{Pruzhinskaya2023}. 
Among them, SNAD160 displays a particularly broad LC and further analysis indicates that it could belong to the pair instability supernovae class \cite{2022RNAAS...6..122P}. It is a theoretical category of thermonuclear explosions of extremely massive stars, as described by \cite{PhysRevLett.18.379} \cite{doi:10.1146/annurev-astro-081817-051819}. There have been no reports of observational confirmation of such event yet and any good candidate contributes to enrich our understanding of the phenomena. \\

Additional analysis has been performed by inputting SNAD160 to a symbolic regression algorithm. It is a machine learning method based on genetic programming \cite{koza1992genetic} that computes a mathematical expression with independent variables to optimally fits some input data. We developed a more complex approach of the algorithm called Multiview symbolic regression (MvSR). It allows for the input of multiple datasets supposedly generated by the same parametric function, and directly returns the common parametric function that generated the examples. MvSR was applied to SNAD160 with LCs in passbands ``g'' and ``r'' being used as two separate datasets. The resulting parametric function describes transient like behaviors with a linear rising and an exponential decay. 

\begin{equation}
    f(t) = A(t+C)\times e^{B(t+C)}
\label{eq:slsn}
\end{equation}

\begin{figure}
    \centering
    \includegraphics[width=1\textwidth]{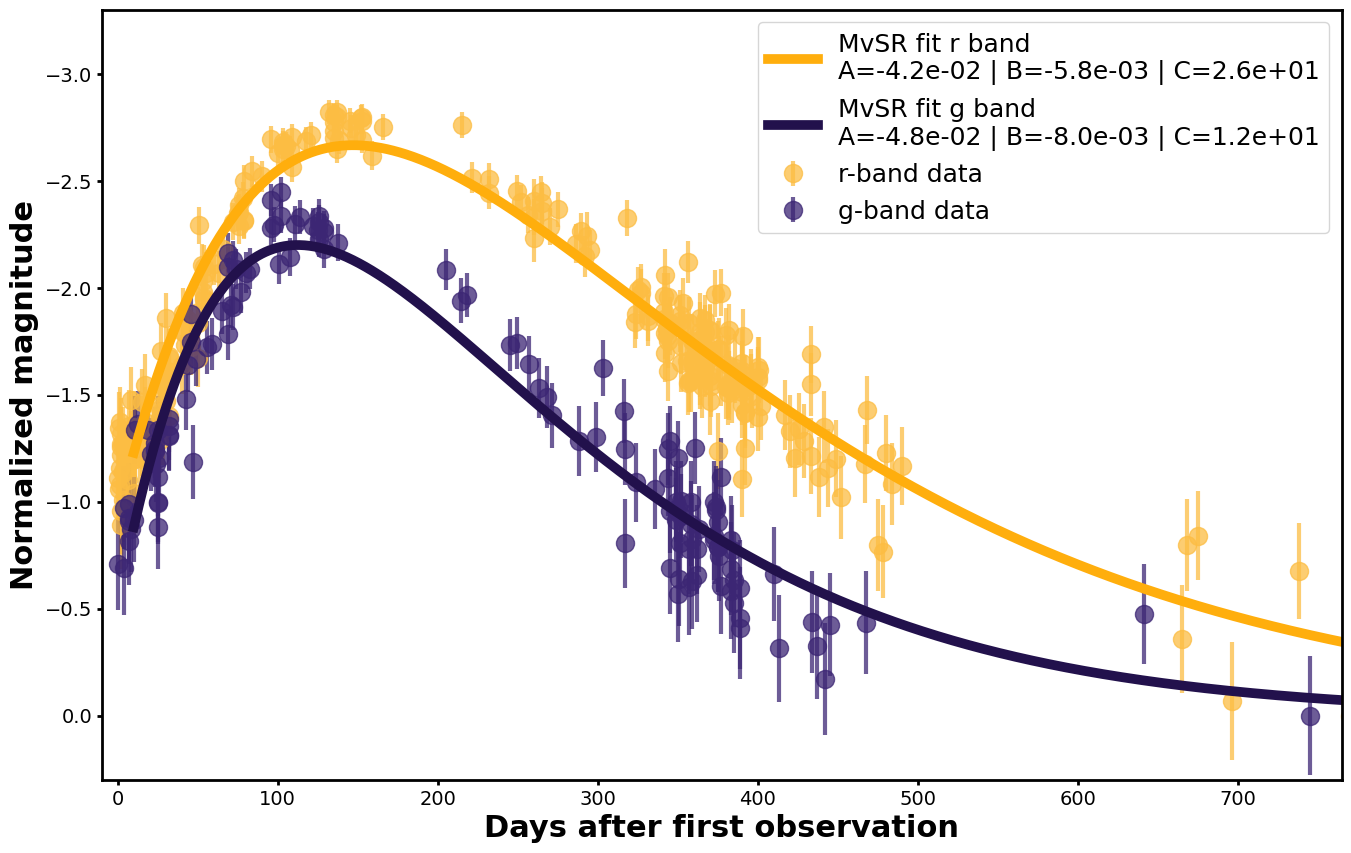}
    \caption{Fit of SNAD160 using equation \ref{eq:slsn} resulting from  MvSR procedure. The magnitude is normalized by subtraction of the peak magnitude.}
    \label{fig:SNAD160}
\end{figure}

It is particularly well fitting SNAD160 in both pass-bands as shown in Fig.~\ref{fig:SNAD160}. It has been applied to other SLSN candidates and provides an accurate representation of the LCs. With three free parameters, this equation offers a noticeably dense description of supernova events. Using this representation for further feature extraction analysis could lead to more interesting SLSN discoveries.

\subsection{Red dwarf flares}


Within the SNAD group, we are currently working on detecting red dwarf flares in the high cadence data of the ZTF survey releases. Stellar flares are very energetic phenomena, during which the optical luminosity of a star increases by several fold over tens of seconds.
The currently accepted version of stellar flares' nature is a magnetic field, generated in the convective stellar interior~\cite{Gershberg}.
The detection of rapid stellar flares, such as those from red dwarfs, is of significant importance in the field of exoplanet science, particularly when studying exoplanet habitability. 
Additionally, creating a statistically significant sample of red dwarf flares can greatly aid in our understanding of the physical processes involved in these events.

By this time, approximately a hundred new candidates for red dwarf flares have been found (A. Voloshina et al., in prep). Some of them were discovered using the AAD algorithm  (Section \ref{subsec:data_models}). Others were detected by fitting the LCs with a model that accurately describes the properties and shape of red dwarf flares' LCs, followed by further analysis of the goodness of fit. The candidates found using presented methods are then subject to further analysis and approval by experts within the SNAD team, with the assistance of the SNAD ZTF Viewer (Section \ref{viewer}).

\begin{figure}
    \centering
    \includegraphics[width=\columnwidth]{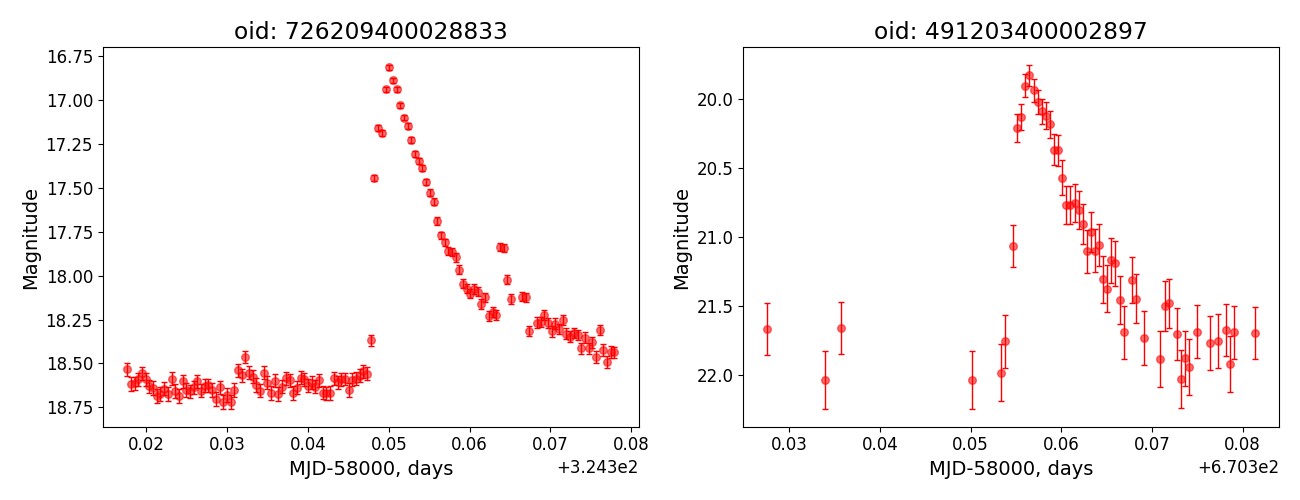}
  \caption{Two examples of red dwarf flares candidates found by SNAD team.}
  \label{fig:rdf_candidates}
\end{figure}

Examples of found red dwarf flare candidates LCs presented in Fig.~\ref{fig:rdf_candidates}.
Found candidates vary both by a LC shape (multiple outburst or one outburst flare) and brightness, which allows exploring the complexity of flare events and forming a more comprehensive sample.

\subsection{Catalog of artefacts}
\label{subsec:artefacts}
\begin{figure*}[ht]
    \centering
    \subfloat{\includegraphics[width=0.2\linewidth]{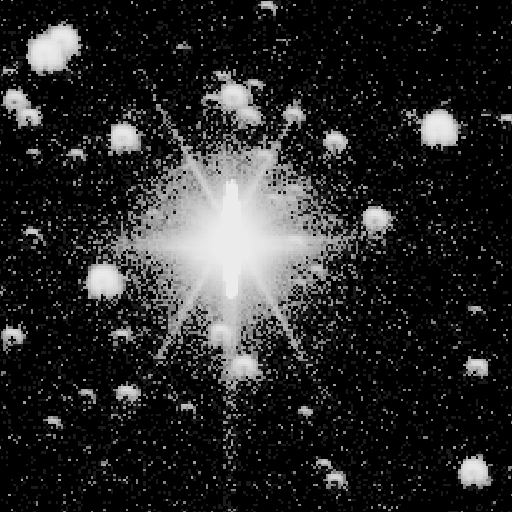}}
    \subfloat{\includegraphics[width=0.2\linewidth]{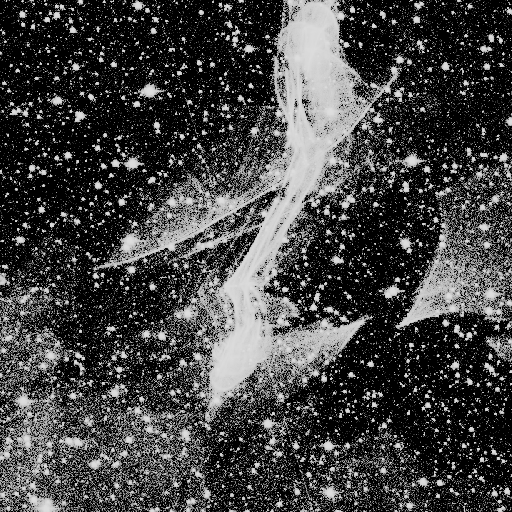}} 
    \subfloat{\includegraphics[width=0.2\linewidth]{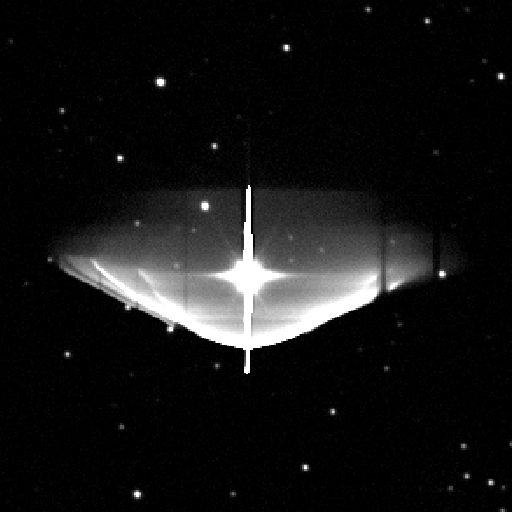}} 
    \subfloat{\includegraphics[width=0.2\linewidth]{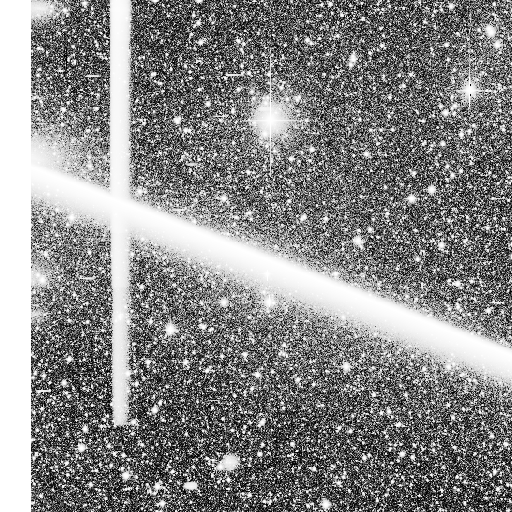}} 
\caption{Some examples from the SNAD catalog of artefacts.}
\label{fig : artefacts}
\end{figure*}

Another interesting data product that resulted from the construction of the SNAD database is the SNAD catalog of artefacts. The term artefact is generally used as an umbrella term to denote observational, phenomenological or instrumental oddities in signal that manifest as diffraction spikes, a step-effect in brightness, colour saturation etc. to name a few. We are collating a catalog of such occurrences in the SNAD database that is aimed for both outreach and science goals. The catalog consists of the OIDs (identifying number that's referenced in ZTF), the FITS images of the respective fields and the SNAD link to the main object page denoted by the OID. Some examples of the artefacts thus identified are given in Figure \ref{fig : artefacts}. While spurious in the strictest sense, several of these artefacts present as visually breathtaking making them perfect for outreach purposes. 

The catalog of artefacts also provides a data set for training machine learning algorithms that could potentially be used for identifying and/or removing these objects from fields of interest. Using the labels of 2230 objects from SNAD knowledge database (Section~\ref{akb}), roughly half of which are artefacts,  we developed an algorithm that  predicts whether an object is an artefact or not  based on the sequence of  object frames from the ZTF survey. The algorithm consists of two parts: (1) a variational auto-encoder that returns the compressed representation of each frame after being trained on all the images from the sample and, (2) a recurrent neural network that takes the sequence of compressed frame representations and an object label as inputs, and returns the probability that the object is an artefact. Currently the best classification result is $\rm{ROC-AUC} = 0.856 \pm 0.010 $  and $\rm{Accuracy} = 0.802 \pm 0.023$.

\section{Scientific Tools and Resources}
\label{sec:public}

\subsection{SNAD ZTF Viewer}
\label{viewer}

The SNAD Viewer\footnote{\url{https://ztf.snad.space}} \cite{ZTFviewer} is a web portal specifically designed for astronomers, providing a centralized view of individual objects from various data releases.
It integrates data from multiple publicly available astronomical archives and sources, thereby offering a comprehensive platform for time-domain data analysis and interpretation.

In the era of big data in astronomy, the SNAD Viewer emerges as a significant tool designed to centralize and streamline the management of astronomical data.
Initially conceived to facilitate expert feedback in active machine learning applications (see Sections~\ref{miner},\ref{subsec:pine}), the SNAD Viewer has evolved into a valuable community asset.
It centralizes public information and provides a multi-dimensional view of individual objects from the ZTF data releases.

The SNAD Viewer's infrastructure is characterized by its scalability and flexibility.
It is designed to accommodate a wide range of data types and user needs, demonstrating its adaptability to the changing landscape of astronomical research.
This adaptability extends to its ability to be personalized and used by other surveys and for various scientific goals, underscoring its broad applicability within the field.

Importantly, the SNAD Viewer is not an isolated entity but is part of a larger network of astronomical data portals.
It is linked to by other significant portals such as Antares, Fink, YSE PZ, and Astro-COLIBRI, which further enhances its accessibility and utility within the astronomical community.
This interconnectedness highlights the collaborative nature of modern astronomical research and the crucial role of the SNAD Viewer within this ecosystem.

The SNAD Viewer is publicly available online, emphasizing its commitment to open science and the democratization of astronomical data.
It serves as a testament to the crucial roles that domain experts continue to play in the era of big data in astronomy.


\subsection{SNAD knowledge database}
\label{akb}
In addition to the discovery of candidates in anomalies, our research also led to the creation of a knowledge database. One of the core features of the SNAD ZTF Viewer (Section~\ref{viewer}) is the ability to assign specific tags to each ZTF object, a function our experts extensively used when reviewing potential anomalies.
This tagging system incorporates a wide range of general astronomical classes, including variable stars, transients, active galactic nuclei, along with their various specific types and subtypes. We also devised custom tags for internal purposes, as well as non-astrophysical tags like artefacts and their subtypes. Multiple tags can be assigned to a single object, with the history of tag changes stored in the database.

During the expert analysis, a total of 2743 objects were labelled. Despite the ZTF data processing pipeline's procedure to distinguish astrophysical events from bogus ones, almost a half of them are artefacts. These assigned labels serve as a reliable source of curated data for training supervised machine learning models. Furthermore, they can help to refine the ZTF's pre-processing pipeline, ensuring more efficient identification and classification of astronomical events in the future.


\subsection{SNAD Transient Miner}
\label{miner}

The SNAD Miner is an exhaustive similarity search for LC features for transient discovery, using simulated LCs as a guide to identify transients in a large dataset mostly comprised of variable stars~\cite{2020ApJS..249...18C}. Specifically, it leverages extracted statistical features from simulated SNe LCs, from which a k-D tree is applied to search for similar feature values in the ZTF~DR4. As a result, we discovered 11 previously missed transients, 7 SNe candidates and 4 AGN candidates.

Realistic ZTF simulations were generated using \texttt{SNANA}~\cite{Kessler2009}, with ZTF data release~3 cadence and magnitude error distribution
(see \cite{Chatterjee2021} for details). Starting from the template models originally developed for PLAsTiCC, \cite{Kessler2019,Hlozek2020},  we selected the seven brightest, well-sampled LCs to use as a reference (3 SLSN-I, 1 SN Ia, 1 SN II and 2 TDE with peak magnitude $\sim$17$^m$).
\par

Using these objects, we applied a k-D tree \cite{Bentley1975} to their extracted 82 non-normalized features. Subsequently, we identified the 15 nearest neighbors for each simulation (105 matches in total, resulting in 89 unique ZTF DR4 sources), and manually inspected the results. Ultimately, we discovered 11 previously unreported supernova and active galactic nucleus candidates. The remaining 94 matches (81 unique ZTF DR4 sources) were either known/already reported transients or variable stars. Given the large number of variable stars previously estimated in ZTF data releases \cite{2020ApJS..249...18C}, it is reasonable to expect many well sampled, high-amplitude variable stars whose coverage in parameter space significantly overlaps with the regions populated by transients (see, e.g. Figure~4 in \cite{Aleo2022}). Thus, a ratio of 18 transients (11 newly-discovered) to 44 variable stars out of 89 unique sources selected from 990,220 considered ZTF sources is a very successful result. Considering the extreme case where $\approx 3000$ SNe discovered by ZTF \cite{dhawan2022} were part of the data set, the expected incidence of SNe when choosing 100 sources at random would be of $< 1 (\approx 0.3)$ event. 
\begin{figure*}
    \centering
    \includegraphics[width=\columnwidth]{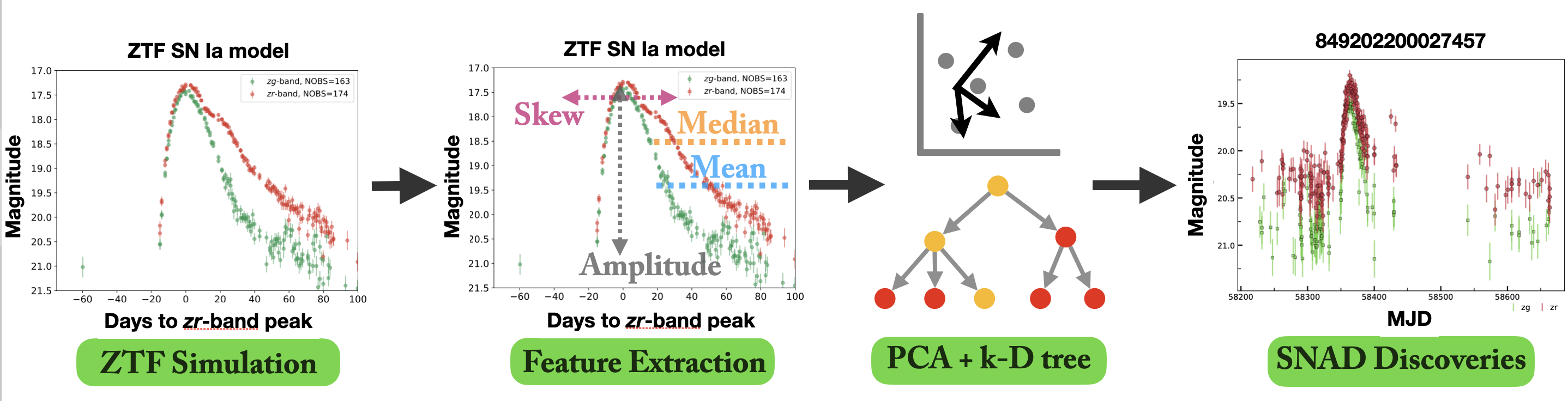}
  \caption{SNAD Miner schematic. We use bright ZTF SNe simulations (left) and extract their LC features (left center). Then, we apply a (PCA+) k-D tree on these features, to search for real ZTF DR events nearest neighbors (right center). Some of these matched nearest neighbors were previously missed SNe (right).}
  \label{fig:miner_schematic}
\end{figure*}

This ``\texttt{SNAD} Miner" process is flexible, and can use simulated or real objects as the input, any number of features, and any number of nearest neighbors to ``mine" in the existing dataset. A schematic is shown in Figure~\ref{fig:miner_schematic}.

\subsection{Coniferest Python Library}
\label{subsec:pine}

The coniferest library \cite{pineforest} is aiming to add adaptive capabilities to isolation forest algorithm which is
inherently static.

The library has implementations of two adaptive learning algorithms. One of them is an already 
mentioned earlier AAD algorithm \cite{Das2017}. And the second one is the implementation of our own adaptive
learning isolation forest algorithm named Pineforest. It is based of tree filtering approach.
After every new observation with some label given by an expert the Pineforest filters out trees that
do not push forward right observations.

This approach have a few remarkable advantages. First, the algorithm have not much hyperparameters, so it
is easy to tune. Second, it maybe used for both -- as an adaptive learning with an expert in loop and
as an accumulator of prior knowledge about data. Finally, it has a very good performance characteristics making it suitable for data intensive applications.

Also, as a bonus the library includes our own implementations of classical
isolation forest with much better performance in scoring compared to scikit-learn's one.
Refer to coniferest package documentation for more details.


\section{Future Directions and Challenges}
\label{future}

Large scale sky surveys have fundamentally changed the process of astronomical investigation and discovery. In the era of LSST, when millions of new transients will be detected every night, serendipitous discoveries will not happen. On the other hand, targeted searches are bound to identify objects which fall within our domain knowledge. 

The SNAD team has been consistently working on the development of adaptable algorithms and tools which allow the user to probe the boundaries of their domain knowledge -- thus enabling scientific discovery in large data sets. These have been proven to be effective in current state of the art catalog data.

In the near future we intend to concentrate our efforts in two important bottlenecks: ensuring that our tools are scalable to meet LSST requirements and make additional connections with expert communities to guarantee scientific impact of SNAD products. The latter will allow the SNAD team to develop increasingly more personalized tools which will fulfill the requirements of individual experts and ensure that we can fully exploit the scientific potential of modern astronomical surveys. 


\subsubsection*{Acknowledgements} Authors thank the Ministry of Science and Higher Education of Russian Federation for financial support, grant
075-15-2022-1221 (2022-BRICS-8847-2335).
We used the equipment funded by the Lomonosov Moscow State University Program of Development.
This work made use of the Illinois Campus Cluster, a computing resource that is operated by the Illinois Campus Cluster Program (ICCP) in conjunction with the National Center for Supercomputing
Applications (NCSA) and which is supported by funds from the University of Illinois at Urbana-Champaign.

%
%
\bibliographystyle{splncs04}
\bibliography{snad-damdid}

\end{document}